# Environmental dependence of different colors in the CMASS sample of the SDSS DR9

Xin-Fa Deng

School of Science, Nanchang University, Jiangxi, China, 330031

**Abstract** In this study, I investigate the environmental dependence of galaxy colors in the CMASS sample of the Sloan Digital Sky Survey Data Release 9 (SDSS DR9). To decrease the radial selection effect, I divide the CMASS sample into subsamples with a redshift binning size of $\Delta z = 0.01$ and analyze the environmental dependence of the u-r, u-g, g-r, r-i and i-z colors for these subsamples in each redshift bin. Statistical analysis shows that all the five colors weakly correlate with the local environment, which may mean that the environmental processes responsible for a galaxy's properties proceed slowly over cosmic time.

**Subject headings** Galaxy: fundamental parameters-- galaxies: statistics

1. **Introduction**

The study of the environmental dependence of galaxy colors has long been an important field (e.g., Brown et al. 2000; Zehavi et al. 2002; Bernardi et al. 2003; Blanton et al. 2003, 2005; Balogh et al. 2004a; Hogg et al. 2004; Lee et al. 2004; Tanaka et al. 2004; Cooper et al. 2006, 2007, 2010; Cucciati et al. 2006; Cassata et al. 2007; Gerke et al. 2007; Bamford et al. 2009; Pannella et al. 2009; Tasca et al. 2009; Iovino et al. 2010; Deng et al. 2007a-b, 2008a-b, 2009a-b, 2010a-c; Skibba et al. 2009; Lee et al. 2010; Wilman et al. 2010; Grützbauch et al. 2011a-b). Numerous studies have focused on the local Universe (e.g., Brown et al. 2000; Zehavi et al. 2002; Bernardi et al. 2003; Blanton et al. 2003, 2005; Balogh et al. 2004a; Hogg et al. 2004; Lee et al. 2004; Tanaka et al. 2004; Deng et al. 2007a-b, 2008a-b, 2009a-b, 2010a-c), and it is widely believed that red galaxies tend to reside in the densest regions of the universe, while blue galaxies tend to reside in lower density regions. The question naturally arises as to whether a strong environmental dependence of galaxy colors might extend to intermediate- and high-redshift regions. This question is a subject of debate. Using the data from the DEEP2 Galaxy Redshift Survey (Davis et al. 2003), Cooper et al. (2006) found that the environmental dependence of galaxy colors at z $\approx$ 1 mirrors that observed in the local Universe. Cooper et al. (2007) also claimed that this strong color–density relation still exists at z>1; however, Cucciati et al. (2006) argued that in the redshift range of 0.25 < z < 0.60, the correlation between the color and local density decreases progressively with increasing redshift until it is undetectable at z $\approx$ 0.9. Grützbauch et al. (2011a) demonstrated that a galaxy's color weakly correlate with the local number density in the redshift range of 0.4 < z < 1. For redshifts up to z $\approx$ 3, Grützbauch et al. (2011b) did not find a strong environmental dependence of galaxy colors.

The Sloan Digital Sky Survey III (SDSS-III; Eisenstein et al. 2011) includes four surveys: SEGUE-2, the Baryon Oscillation Spectroscopic Survey (BOSS), the Multi-object APO Radial Velocity Exoplanet Large-area Survey (MARVELS) and the Apache Point Observatory Galactic Evolution Experiment (APOGEE). The primary goal of the BOSS project is to carry out a redshift survey of 1.5 million luminous red galaxies (LRGs) at 0.15 < z < 0.8 over 10, 000 square degrees and 160,000 quasi-stellar objects (QSOs) at 2.15 < z < 3.5 over 8,000 square degrees.

Undoubtedly, the BOSS galaxy sample is a valuable sample in dealing with intermediate redshifts. Here, I explore the environmental dependence of colors in this sample to further understand the color-density relation in the intermediate redshift regime.

The outline of this paper is as follows. In Section 2, I describe the data used. I describe the statistical method in Section 3. In Section 4, I discuss the environmental dependence of colors in the BOSS galaxy sample. I summarize my main results and conclusions in Section 5.

In calculating the distance, I used a cosmological model with a matter density $\Omega_0 = 0.3$, cosmological constant $\Omega_\Lambda = 0.7$, and Hubble's constant $H_0 = 70 \text{km} \cdot \text{s}^{-1} \cdot \text{Mpc}^{-1}$.

## 2. Data

In this study, I used the galaxy data from the ninth data release (DR9) (Ahn et al. 2012) of the SDSS, which is the first public release of spectroscopic data from the SDSS-III Baryon Oscillation Spectroscopic Survey (BOSS). DR9 includes 535,995 new galaxy spectra (median z $\approx$ 0.52), 102,100 new quasar spectra (median z $\approx$ 2.32), and 90,897 new stellar spectra, along with the data presented in previous data releases.

The BOSS galaxy sample is divided into two principal samples at z $\approx$ 0.4: "LOWZ" and "CMASS". The LOWZ sample is a simple extension of the SDSS-I and –II LRG sample (Eisenstein et al. 2001), which is a low redshift sample with a median redshift of z = 0.3. The majority of galaxies in the LOWZ sample are located in the redshift range of 0.15 < z < 0.43. The CMASS sample is designed to select galaxies above z $\approx$ 0.4 and is a nearly complete sample of massive galaxies above the magnitude limit of the survey that have intermediate redshifts, which represents a probe of an entirely new cosmological volume. I therefore restrict this study to the CMASS sample.

The data was downloaded from the Catalog Archive Server of SDSS Data Release 9 (Ahn et al. 2012) by the SDSS SQL Search (http://www.sdss3.org/dr9/). Because most CMASS galaxies are located between 0.43 < z < 0.7, I extracted 296501 CMASS galaxies (with SDSS flag: BOSS_TARGET1&128>0) in the redshift region $0.43 \leq z \leq 0.7$ with stellar masses calculated by Maraston et al. (2013) (http://data.sdss3.org/dr9/boss/spectro/redux/galaxy/). Maraston et al. (2013) employed two template fittings (passive and star-forming ) and two adopted Initial Mass Functions (IMFs) (Salpeter and Kroupa ); they also considered the mass lost via stellar evolution. The passive model does not include the possibility of a non-zero SFR (star formation rate). The selection of the star-forming template and the Kroupa IMF leads to the largest number of non-zero SFR galaxies. Considering that further investigation would likely shed light on the SFR of galaxies, I used the data of best-fit stellar mass [in log $M_{sun}$] obtained with the star-forming template and the Kroupa IMF.

In this study, the model magnitudes are used. In all cases, galactic extinction corrections are applied, but k-corrections are not used. Maraston et al. (2013) indicated that BOSS is a mass-uniform sample over the redshift range 0.2 to 0.6. Dawson et al. (2013) argued that the BOSS galaxies have an approximately uniform co-moving number density out to a redshift of z = 0.6, but at z $\approx$ 0.8, the co-moving number density of BOSS galaxies decreases monotonically to zero. Fig.2 of Anderson et al. (2012) also demonstrated the number-density of CMASS galaxies drops dramatically with increasing redshift at z>0.6. Because the radial selection effect of the CMASS sample is fairly serious at redshift z>0.6, I limited the sampling to 212911 CMASS galaxies with a redshift of $0.44 \leq z \leq 0.59$ and constructed a mass-uniform sample.

## 3. Statistical method

Although the CMASS galaxy sample in this work is limited to the range $0.44 \leq z \leq 0.59$, the radial selection effect still exists. To avoid this bias, one often constructs volume-limited samples from the apparent-magnitude limited sample; however, it is difficult to construct an ideal volume-limited sample from the CMASS galaxy sample because it is not a simple flux-limited sample. In the CMASS galaxy sample, the radial selection function is very complicated.

As indicated by Deng (2012), the use of volume-limited samples results in a large fraction of the data becoming useless. With this in mind, Deng (2012) analyzed the apparent-magnitude limited sample to make the maximum use of observational data; however, it is important to remember that the radial selection effect in the apparent-magnitude limited sample is serious. To decrease the radial selection effect, Deng (2012) divided the entire apparent-magnitude limited Main galaxy sample (Strauss et al. 2002) into subsamples with a redshift binning size of $\Delta z = 0.01$ and focused on a statistical analysis of the subsamples in each redshift bin. Following Deng (2012), I measured the projected local density $\sum_5$, which is computed from the distance to the 5th nearest neighbor within a redshift slice $\pm 1000$ km s$^{-1}$ of each galaxy (e.g., Goto et al. 2003; Balogh et al. 2004a, 2004b), divided the CMASS galaxy sample into subsamples with a redshift binning size of $\Delta z = 0.01$, and finally analyzed the environmental dependence of colors for the subsamples in each redshift bin.

Like Deng et al. (2008a), I arranged the galaxies in order from the smallest to the largest for each subsample, selected approximately 5% of the galaxies, constructed two samples at both density extremes, and compared the distributions of galaxy colors in the lowest density regime with those in the highest density regime.

Deng (2012) argued that in each subsample with a redshift binning size of $\Delta z = 0.01$, the radial selection effect and K-corrections are less important and can be ignored. Generally, the K-corrections should be applied when studying colors of galaxies; however, due to the lack of knowledge of the SEDs (spectral energy distributions), K-corrections have inherent uncertainties. There is a gap between the $g$- and $r$-bands, which allows considerable freedom in fitting the data when reconstructing SEDs. Deng (2012) indicated that the application of the K-corrections will produce new subjective biases or assumptions. In fact, Blanton et al. (2003) emphasized that K-corrections are not always necessary or appropriate. In some cases, the use of observed color (without applying K-corrections) is also a reasonable choice.

## 4. Environmental dependence of different colors in the CMASS galaxy sample

Deng (2012) plotted $u$-, $g$-, $r$-, $i$- and $z$-band absolute magnitude distributions at both density extremes in different redshift bins for the apparent-magnitude limited Main galaxy sample of the SDSS DR7 (Abazajian et al. 2009) and found that in each redshift bin, the luminosities of the subsamples in all five passbands apparently correlate with the local environment. Following Deng (2012), Deng et al. (2012) investigated the environmental dependence of the stellar mass, star formation rate (SFR), specific star formation rate (SSFR, the star formation rate per unit stellar mass) and active galactic nucleus (AGN) activity and demonstrated that there is a strong environmental dependence of stellar mass, SFR and SSFR in nearly all redshift bins. They also argued that in most redshift bins (except low redshift region $0.02 \leq z \leq 0.06$), the fraction of AGNs in the sample at low density apparently is larger than that in the sample at high density. Using the same statistical method and galaxy sample, Deng et al. (2013) found that the u-r, u-g, g-r, r-i and i-z colors of galaxies strongly correlate with the local environment in the redshift region $0.05 \leq z \leq 0.14$: red galaxies tend to be located in high density regions, while blue galaxies tend to be located in low density regions. These studies indicate that using a redshift bin of $\Delta z = 0.01$, the environmental dependence of galaxy properties can still be observed, if it exists.

Figs.1-5 show the u-r, u-g, g-r, r-i and i-z color distributions at both density extremes in the different redshift bins for the CMASS galaxy sample. As shown by these figures, all the five colors show minimal correlation with the local environment.

Following Deng (2012) and Deng et al. (2012), I also perform the Kolmogorov-Smirnov (KS) test, which checks whether two independent distributions are similar by calculating a probability value. A large probability implies that it is very likely that the two distributions are derived from the same parent distribution. Conversely, a low probability implies that the two distributions are different. The probability that the two distributions come from the same parent distribution is listed in Table 1, which is much larger than that in the apparent-magnitude limited Main galaxy sample (see Table 1 of Deng 2012 and Deng et al. 2012) and is much larger than 0.05 (5% being the standard in a statistical analysis) in many redshift bins. This result is in good agreement with the conclusion obtained by the step figures.

Grützbauch et al. (2011a) demonstrated that galaxy color very strongly correlate with stellar mass at $0.4 < z < 1$, and has a weak environmental dependence only at lower redshifts ($0.4 < z < 0.7$). They also found a weak stellar mass dependence on the environment at intermediate redshifts and claimed that the color-density relation is a combination of a strong color-stellar mass relation and a weak stellar mass-density relation. Grützbauch et al. (2011a) observed that the environmental influence of galaxy colors is clearest in intermediate mass galaxies ($10.5 < \log M_* < 11$), whereas colors of lower and higher mass galaxies are insensitive to their redshift and environment. Fig.6 shows the stellar mass distribution for the CMASS sample with redshifts of $0.44 \leq z \leq 0.59$. As shown by this figure, there is a fairly high fraction of galaxies with higher stellar masses in this sample. The percentage of intermediate mass galaxies ($10.5 < \log M_* < 11$) is only 30.08. Thus, it is not surprising that colors of CMASS galaxies show a weak dependence on the environment.

Grützbauch et al. (2011b) further showed that the colors of galaxies are strongly dependent on the stellar mass at redshifts up to $z \approx 3$ and argued that stellar mass is the most important factor in determining the colors of galaxies in the early universe up to $z \approx 3$ and that the local density likely has a small additional effect but only at the most extreme overdensities. As indicated by Grützbauch et al. (2011b), a possible interpretation for this is that the environmental processes that exert the essential influence on galaxy properties proceed slowly over cosmic time. Some of the most influential high-density environments may still be in the process of being built up and cannot yet affect galaxy colors.

The galaxy sample selection may also likely lead to different environmental dependence of galaxy colors. Grützbauch et al. (2011a) argued that the color difference largely disappears when stellar mass selected samples are used. Cooper et al. (2007) found that a strong relation between color and local density persists out to $z > 1$. Grützbauch et al. (2011a) indicated that this might be partly caused by their sample selection, which is rest-frame B-band luminosity limited. The CMASS sample is not restricted to a sample of red galaxies, but instead attempts to select a stellar mass-limited sample of objects of all intrinsic colors. This work shows that for such a massive and predominantly bulge-dominated sample, the environmental dependence of galaxy colors is fairly weak.

5. Summary

The primary goal of this study is to investigate the environmental dependence of colors in the CMASS sample of the SDSS DR9 (Ahn et al. 2012). Considering that the number-density of CMASS galaxies drops dramatically with increasing redshift at a redshift of $z>0.6$, I restrict my statistical analysis here to the CMASS sample with redshifts of $0.44 \leq z \leq 0.59$, which contains 212911 CMASS galaxies. Following Deng (2012), to decrease the radial selection effect, I divide the CMASS sample into subsamples with a redshift binning size of $\Delta z = 0.01$ and analyze the environmental dependence of u-r, u-g, g-r, r-i and i-z colors for these subsamples in each redshift bin. As shown by Figs.1-5, overall, all the five colors very weakly correlate with the local environment. As indicated by Grützbauch et al. (2011b), a possible interpretation for this is that the environmental processes that exert the essential influence on galaxy properties proceed slowly over cosmic time.

**Acknowledgements**

I thank the anonymous referee for many useful comments and suggestions. This study was supported by the National Natural Science Foundation of China (NSFC, Grant 11263005). The use of the SDSS-III database (http://www.sdss3.org) is acknowledged.

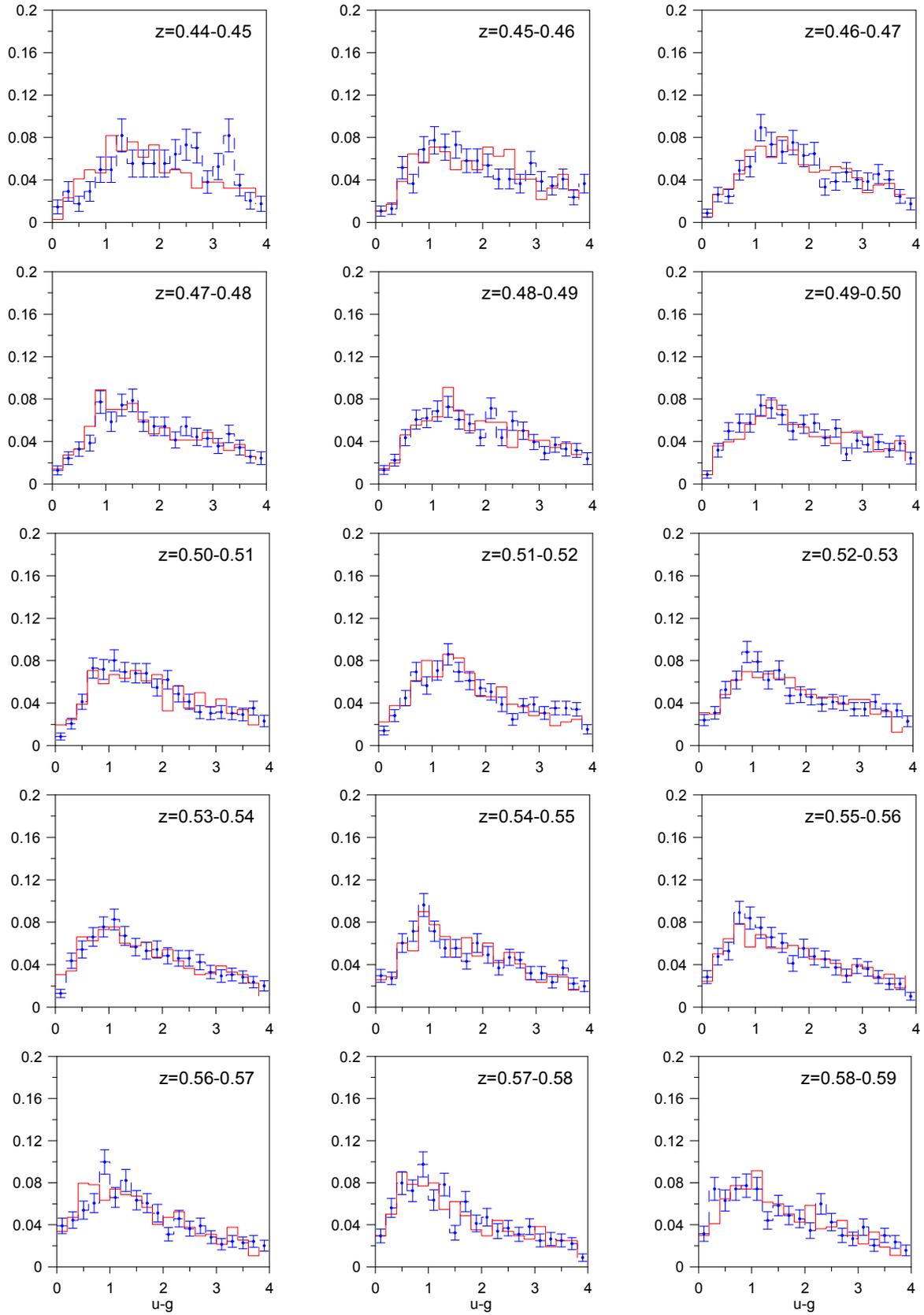

Fig.1 U-g color distribution at both extremes of density in different redshift bins: red solid line for the sample at high density, blue dashed line for the sample at low density. The error bars of blue lines are 1 $\sigma$ Poissonian errors. Error-bars of red lines are omitted for clarity.

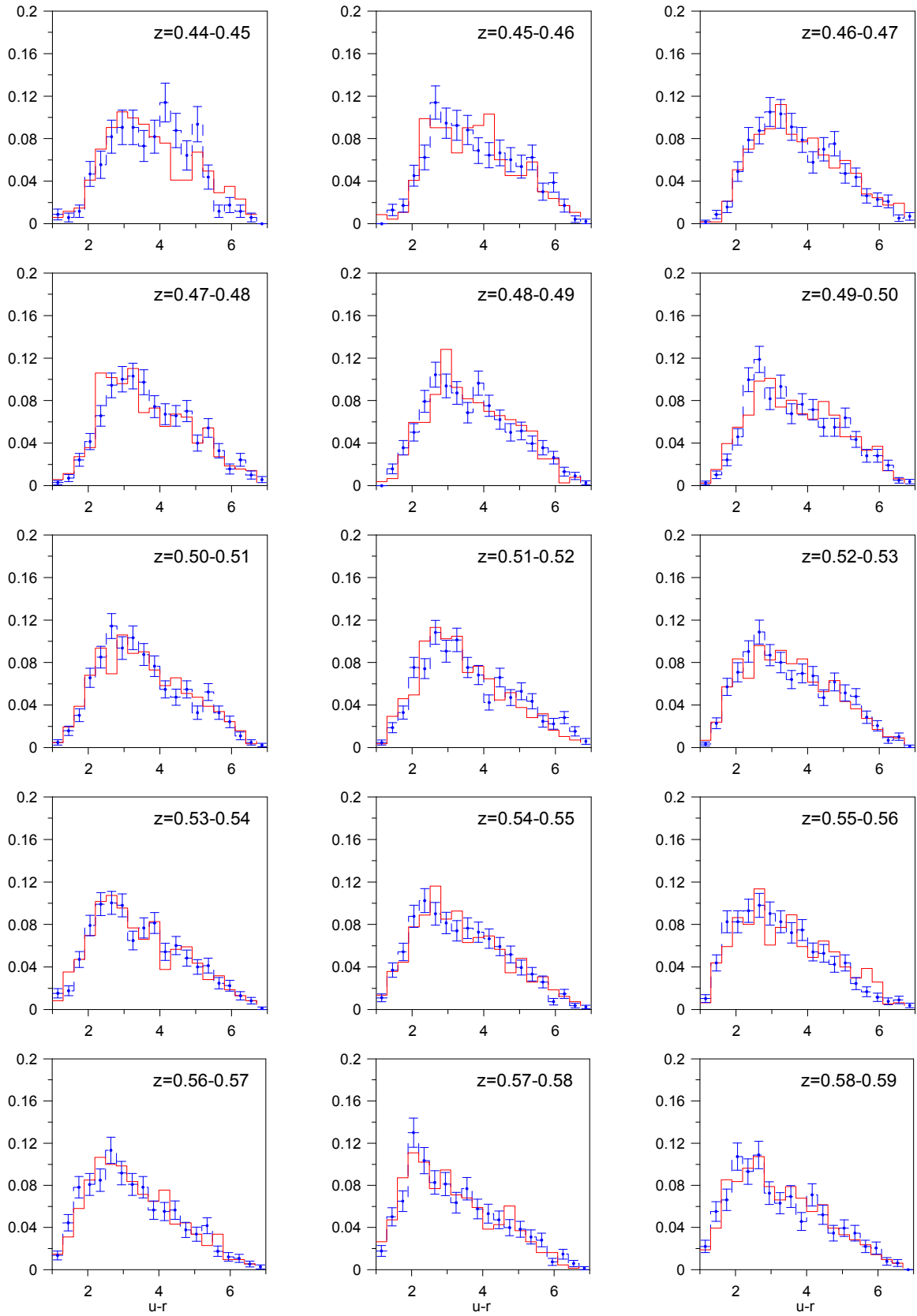

Fig.2 As Fig.1 but for u-r color distribution at both extremes of density in different redshift bins.

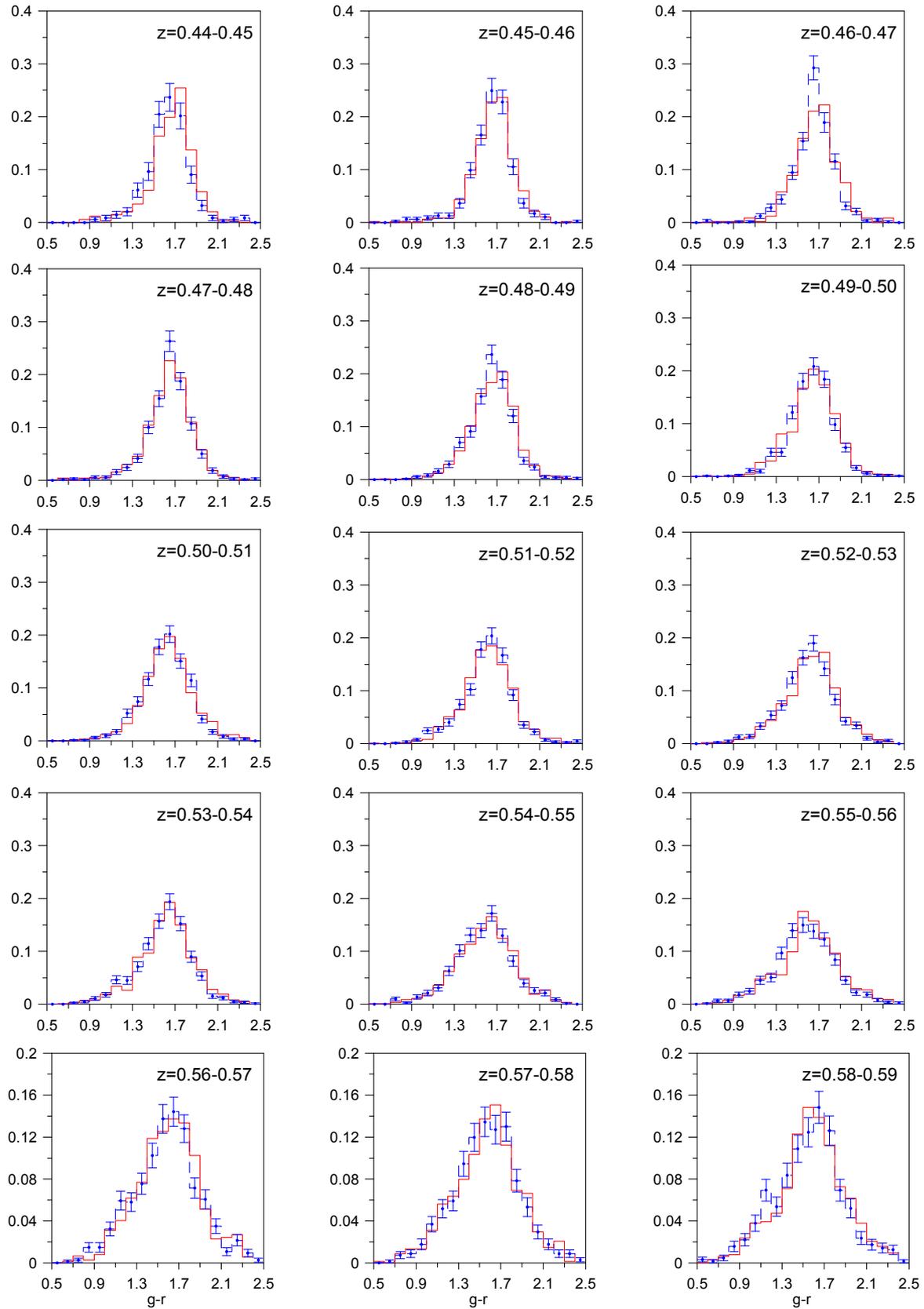

Fig.3 As Fig.1 but for g-r color distribution at both extremes of density in different redshift bins.

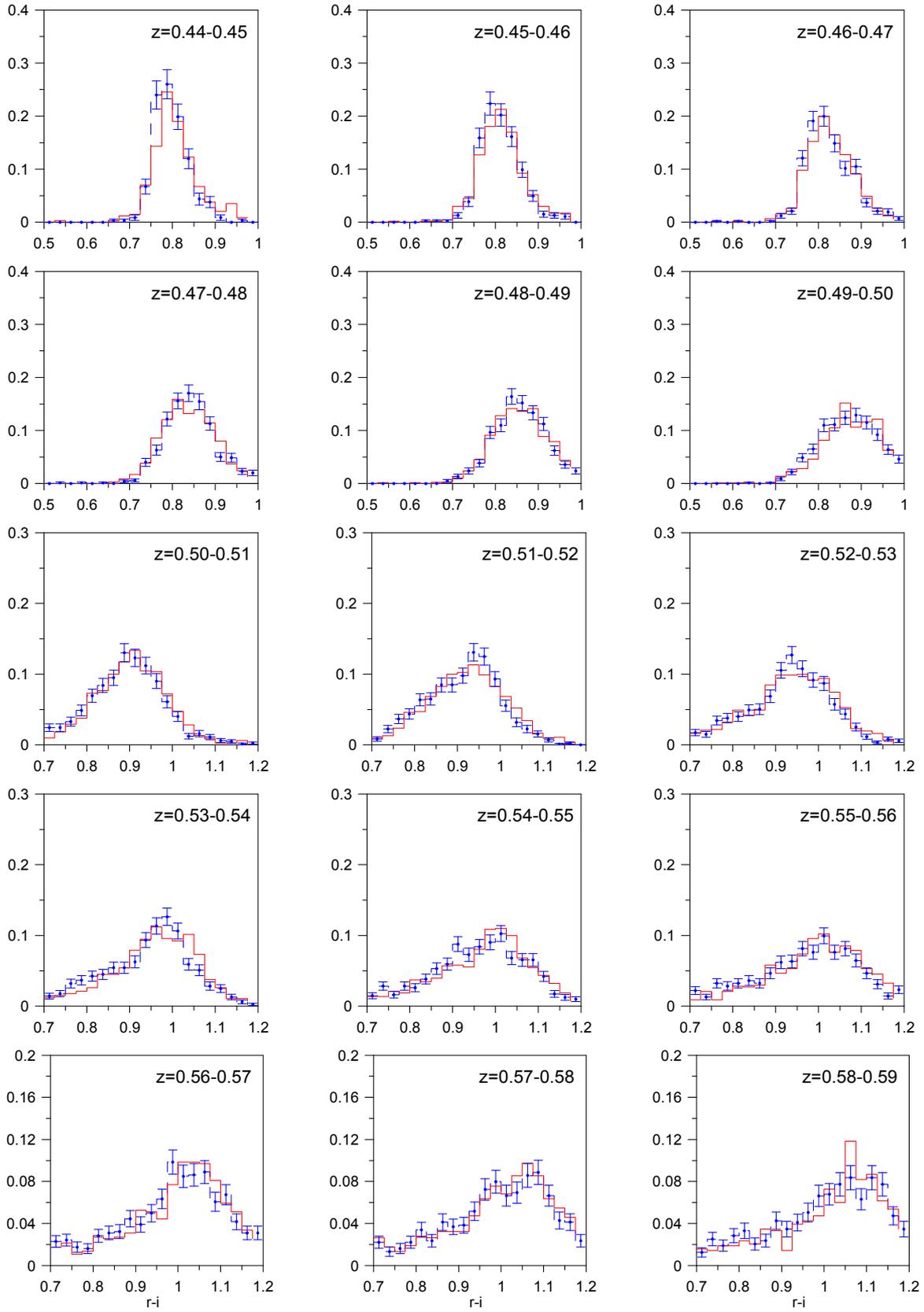

Fig.4 As Fig.1 but for r-i color distribution at both extremes of density in different redshift bins.

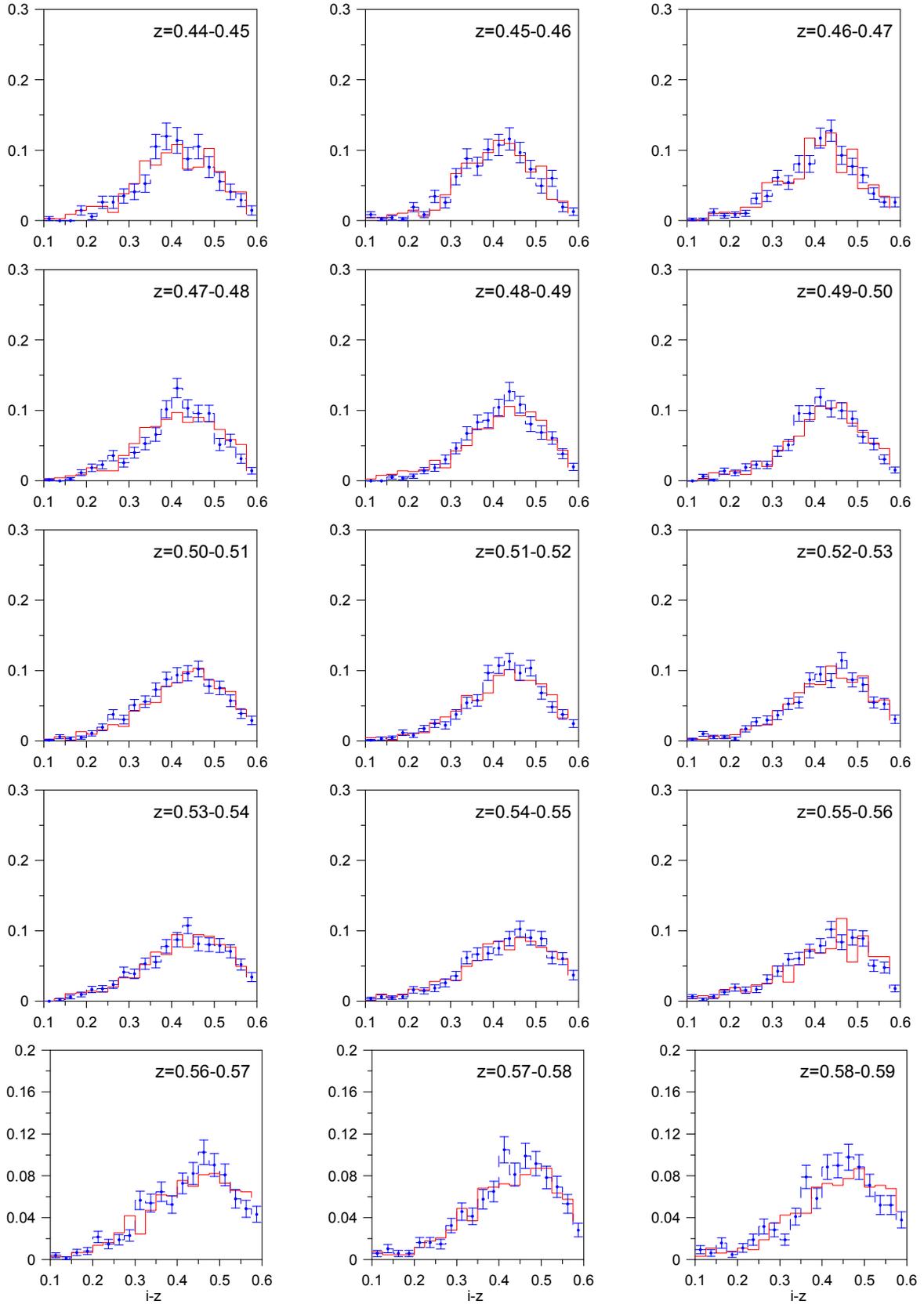

Fig.5 As Fig.1 but for i-z color distribution at both extremes of density in different redshift bins.

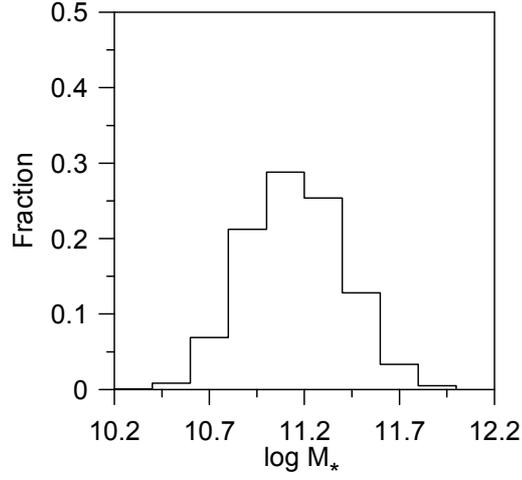

Fig.6 Stellar mass distribution for the CMASS sample with the redshift $0.44 \leq z \leq 0.59$.

Table 1: K–S probabilities of different colors that two samples at both extremes of density are drawn from the same distribution.

| Redshift bins | Galaxy number | Projected local density range (Galaxies Mpc$^{-2}$) | P(u-r) | P(u-g) | P(g-r) | P(r-i) | P(i-z) |
|---|---|---|---|---|---|---|---|
| 0.44-0.45 | 6833 | $1.39\times10^{-4}$—54.57 | 0.163 | 0.0518 | 0.00116 | 0.00208 | 0.655 |
| 0.45-0.46 | 9291 | $1.87\times10^{-4}$—90.17 | 0.610 | 0.721 | 0.358 | 0.451 | 0.999 |
| 0.46-0.47 | 11420 | $1.60\times10^{-4}$—235.25 | 0.443 | 0.828 | 0.000539 | 0.134 | 0.685 |
| 0.47-0.48 | 13970 | $2.84\times10^{-4}$—274.39 | 0.137 | 0.304 | 0.934 | 0.246 | 0.196 |
| 0.48-0.49 | 15146 | $2.22\times10^{-4}$—87.55 | 0.460 | 0.969 | 0.0404 | 0.627 | 0.0222 |
| 0.49-0.50 | 15650 | $2.80\times10^{-4}$—295.00 | 0.563 | 0.814 | 0.227 | 0.0139 | 0.183 |
| 0.50-0.51 | 16444 | $2.32\times10^{-4}$—451.68 | 0.595 | 0.475 | 0.252 | 0.336 | 0.00282 |
| 0.51-0.52 | 16984 | $1.19\times10^{-4}$—1178.79 | 0.0204 | 0.0204 | 0.496 | 0.0416 | 0.144 |
| 0.52-0.53 | 17475 | $3.43\times10^{-4}$—137.17 | 0.793 | 0.554 | 0.00687 | 0.139 | 0.793 |
| 0.53-0.54 | 16938 | $2.18\times10^{-4}$—252.65 | 0.910 | 0.655 | 0.268 | 0.00408 | 0.295 |
| 0.54-0.55 | 16204 | $1.46\times10^{-4}$—102.56 | 0.711 | 0.792 | 0.178 | 0.00618 | 0.830 |
| 0.55-0.56 | 15491 | $1.96\times10^{-4}$—112.18 | 0.111 | 0.178 | 0.00275 | 0.0132 | 0.0132 |
| 0.56-0.57 | 14834 | $1.99\times10^{-4}$—363.07 | 0.614 | 0.787 | 0.373 | 0.0373 | 0.0754 |
| 0.57-0.58 | 13546 | $2.49\times10^{-4}$—186.67 | 0.782 | 0.427 | 0.692 | 0.555 | 0.318 |
| 0.58-0.59 | 12685 | $1.79\times10^{-4}$—71.93 | 0.512 | 0.791 | 0.119 | 0.0312 | 0.00920 |